# X-ray driven first-order phase transition in GeO$_2$ glass under pressure

Xinguo Hong[1,2,*], Matt Newville[3], Ho-Kwang Mao[1,4], T. Irifune[5] and L. Huang[6]

[1]*Center for High Pressure Science and Technology Advanced Research, Beijing 100094, P.R. China*

[2]*Mineral Physics Institute, Stony Brook University, Stony Brook, NY 11794*

[3]*Consortium for Advanced Radiation Sources, University of Chicago, Chicago, Illinois 60637*

[4]*Geophysical Laboratory, Carnegie Institution of Washington, Washington, DC 20015, USA*

[5]*Geodynamics Research Center, Ehime University, Matsuyama 790-8577, Japan*

[6]*Department of Materials Science and Engineering, Rensselaer Polytechnic Institute, Troy, NY 12180, USA*

ABSTRACT

Unprecedented bright and intense synchrotron X-rays have been widely used to unravel numerous compelling electronic and structural properties using a large variety of physical techniques, while the puzzling phenomenon is the uncertainties of measurements due to X-ray induced considerable electronic and structural changes of a matter under study. Here, we report an X-ray driven first-order tetrahedral-octahedral phase transition in GeO$_2$ glass at high pressure using X-ray absorption fine structure (XAFS) with a nano-polycrystalline diamond anvil (NPD) cell. Upon X-ray irradiation from an undulator device, the XAFS spectrum at 5.4 GPa, which is below the threshold pressure of tetrahedral to pentahedral transition, starts to progressively approach the spectrum of octahedral GeO$_2$ at 20.4 GPa. Detailed analysis indicates that both the nearest Ge–O distance and coordination number (CN) in the first Ge–O shell of $GeO_2$ glass at 5.4 GPa increase to those of a fully octahedral glass above 20 GPa, while negligible changes were found at other low pressures. These observations demonstrate that X-ray irradiation can be served as an alternative stimulus for forming dense matter under pressure, as the well-known external stimuli of pressure and temperature. Dense matter formed under external stimuli is crucial for understanding the formation, differentiation and evolution of planet and Earth. Attention for the X-ray induced structural uncertainties is called, especially for addressing metastable states under extreme conditions which may become undetectable in terms of intense synchrotron X-rays as structural tools.

*) Corresponding author

I. Introduction

Knowledge of the interaction between X-ray and matter is a pillar of modern science and technology. The world-wide efforts to improve X-ray flux, brightness and microfocusing at the third/fourth generation synchrotron sources, which have been driven by small samples, local areas and high spatial resolution, are opening unprecedented fascinating opportunities for investigating the microstructure, elemental distribution, and chemical bonding state of advanced materials and biological samples[1-3]. One puzzling phenomenon for structural determination is that intense X-ray itself may induce significant changes in the electronic and structural properties of a matter under study [4-6]. Over the past decades, numerous second-order phase transitions triggered by light or X-ray irradiation, the so-called photoinduced phase transitions, have been reported in a variety of systems, e.g., photo-magnets [7-13], photo-conductivity [4,9,14-24], and spin-crossover complexes[6,13,25-29]. However, a first-order phase transition induced by X-ray irradiation is remarkable and reports are sparse [30].

Exploration of the effects of external perturbations (pressure, temperature, magnetic and electric fields, light or X-ray irradiations) on the structural, electronic, and magnetic properties of materials is a key area of research in contemporary condensed-matter physics, chemistry, geoscience, and materials sciences. Pressure is widely served as a clean external stimulus without introducing chemical impurity. The high-pressure behavior of prototypical network-forming glasses, silica ($SiO_2$) and germania ($GeO_2$) glasses have been extensively studied [31-44]. Previous *in situ* studies on $GeO_2$ glass have revealed that not only the pressure [31,32,36,41,45,46] but also thermal stimulus [43] can lead to an evolution from a continuous random network of corner shared tetrahedra to a dense octahedrally coordinated glass. As pressure increases, multiple polyamorphs of $GeO_2$ glass under high pressure have been unraveled [31,32,35,36,39,40,42,44,46-50].

As an element sensitive synchrotron-based technique, XAFS is a powerful tool to study the structure of local (SRO) and intermediate range order (IRO) surrounding selected chemical species in condensed matter [51]. Despite its complementarity with x-ray diffraction, however, for high pressure research, it has long been regarded that the conventional XAFS technique is unsuitable for the environment of diamond anvil cell (DAC) due to the Bragg reflections from the single crystal diamond anvils (glitches) [52]. Several methods have been proposed to eliminate DAC glitches for high-pressure XAFS measurements [53-56]. An iterative method was proposed to obtain glitch-free XAFS data from single crystal diamond anvils [53], with which *in situ* high-pressure XAFS data have been successfully measured up to 64 GPa [35]. Nevertheless, data acquiring over a set of multiple angles is relatively time consuming. Recently, nano-polycrystalline diamond (NPD) anvil, which consists of randomly oriented nano-grained diamonds, is proven to be promising for glitch-free XAFS spectra [55,57,58]. NPD anvils have been used for a study of $GeO_2$ glass up to 44 GPa [39].

Here, we report an X-ray driven tetrahedral-octahedral transition in $GeO_2$ glass using XAFS technique with a nano-polycrystalline diamond (NPD) anvil cell. XAFS was employed to simultaneously probe the local coordination change triggered by X-ray irradiation. To the best of our knowledge, there is no report on this unusual X-ray triggered amorphous-to-amorphous phase transition in network-forming glass at high pressure so far.

II. Experiment

$GeO_2$ glass was prepared from 99.98% purity germanium dioxide powder, which was melted at 1600 °C in a platinum crucible and held at this temperature for 1 h. The crucible with the melt was quenched from 1600 °C in cold water. The glass sample was annealed at 550 °C for 2 h and furnace cooled to remove residual stress. Princeton-manufactured large–opening symmetrical diamond-anvil cell (DAC) were employed with a pair of 300 μm culet NPD anvils. The sample was loaded in a 120 μm sample chamber of Re gasket without hydrostatic pressure medium. The pressure was measured using the standard ruby fluorescence technique [59].

XAFS experiments were carried out on the Ge $K$-edge of the glass in transmission mode at the GeoSoilEnviroCARS undulator beamline 13-ID-E, Advanced Photon Source (APS), Argonne National Laboratory. The X-ray beam size was focused down both in the horizontal and in the vertical planes to a beam spot size of 1–2 μm using Kirkpatrick-Baez mirrors, spatially stable over the entire energy scan. XAFS spectra were collected from 11,003 to 11,850 eV with 5 eV steps before the main edge, 0.2 eV steps from -10 eV to 25 eV across the main edge (11,103 eV), and 0.05 Å$^{-1}$ steps in $k$-space to 14 Å$^{-1}$ above the main edge. XAFS data processing and analysis were then performed with the ATHENA and ARTEMIS programs [60] of the IFEFFIT package [61].

III. Results and Discussion

Fig. 1 shows the X-ray irradiation effects on the Ge $K$-edge XAFS of GeO$_2$ glass at 2.7 GPa and 5.4 GPa, respectively, which were collected at the undulator beamline13-ID-E. The quality of XAFS raw data is excellent, featuring no obvious spurious glitch from the NPD anvils. It is notable that the overall spectra at 5.4 GPa (Left panel, Fig. 1) changes significantly as irradiation time increases while the data at 2.7 GPa remain nearly the same for all three different measurements.

In the expanded range of X-ray absorption near edge structure (XANES), a dramatic change occurs at 5.4 GPa (Right panel, Fig. 1). This is quite unusual because the GeO$_2$ glass at pressures below 6 GPa remains in tetrahedral form [31,32,45], where previous XAFS measurements repeatedly show only small difference [35,40,54]. The time dependent spectra at 5.4 GPa is unexpected and not understood. Furthermore, with continuous X-ray exposure there is much smaller change for the data collected at 30 min and 45 min, in comparison with that of 15 min. This observation illustrates that X-ray induced structural changes in GeO$_2$ glass becomes terminating, a completion sign of a phase transition.

The enhancement of absorption edge step (Left Panel, inset of Fig. 1a) is proportional to density variation according to absorption law because change of sample thickness at the same pressure is negligible [62]. This implies a densification process of GeO$_2$ glass, which has been triggered by X-ray irradiation. Density is a key physical quantity for distinguishing emerging polyamorphs of non-crystalline materials under external stimuli [35,63].

As the energy of X-ray photons is high enough to cause ionizations of valence electrons, radiation damage occurs with irreversible chemical changes. Synchrotron radiation damage may become problematic since it interferes with the process of phase transition under extreme conditions. However, the similarity of three XAFS spectra measured at 2.7 GPa suggests that the effect of radiation damage is small and should not be responsible for the phenomenon observed at 5.4 GPa.

The transmission of x-ray photons ($I_t$) is equal to the number of x-ray photons shone on the sample ($I_0$) multiplied by a decreasing exponential that depends on the absorption coefficient μ, and the thickness of the sample $x$.

$$I_t = I_0 e^{-\mu x} \qquad (1)$$

The ratio of transmitted photons at the peak of white line ($I_w$) over certain higher energy away, e.g. 100 eV ($I_{w+100}$), is given by,

$$\frac{I_w}{I_{w+100}} = e^{-(\mu x)_w + (\mu x)_{w+100}}$$

$$\cong e^{-5} = 0.67\% \qquad (2)$$

Eq. 2 illustrates that more than two orders of X-ray photons was absorbed, causing a lot of X-ray photoexcitation in the vicinity of white line in contrast to the extended higher extended energy range (EXAFS). The X-ray induced structural change should mostly take place at around near absorption edge, as evidenced by the significant enhancements at white line and XANES range (Fig.1b).

It is informative to compare the obtained spectra with the published data to identify the final state upon X-ray irradiation at 5.4 GPa, i.e., either in $GeO_4$ tetrahedral or in $GeO_6$ octahedral state. Fig. 2 shows a direct comparison with the spectra of tetrahedral glass at 4.6 GPa and octahedral glass at 20.4 GPa, which were measured using a much weaker X-ray beam emitted from a bending magnet source at 13-BM-D, APS [35,53]. The flux (photons/sec) of APS undulator beamline (13-ID-E) is $6 \times 10^{12}$ @ 10 keV while that of 13-BM-D is $1 \times 10^9$ @10 keV. Furthermore, the beam size of focused X-ray becomes one order smaller, i.e., from 15 μm down to 1–2 μm, resulting in roughly $6 \times 10^5$ folds achievement in X-ray brightness at 13-ID-E than that of 13-BM-D.

Fig. 2a shows a comparison of normalized Ge $K$-edge XANES spectra measured at 5.4 GPa (13-ID-E), 4.6 GPa (13-BM-D) and 20.4 GPa (13-BM-D). The XANES profile of X-ray-induced phase measured at 5.4 GPa (13-ID-E) is very similar to that of octahedral glass at 20.4 GPa rather than the tetrahedral one at 4.6 GPa (13-BM-D) despite of large pressure difference. Fig. 2b shows other prominent similarity in the overall $k^2$-weighted XAFS spectra, $k^2\chi(k)$, in k-space. Similar asymmetric oscillation at 6–8 Å$^{-1}$ is observed for octahedron (20.4 GPa) and X-ray modified spectra (5.4 GPa). The corresponding XAFS Fourier transform, $|\chi(R)|$, in $r$-space and the back-transformed $|\chi(R)|$ at 1–3 Å in $q$-space are presented in Fig. 2c and Fig. 2d, respectively. All these observations in XAFS amplitudes and frequencies clearly indicate an octahedral glass formed at a much lower pressure than that of fully octahedral state which has been defined over a wide pressure range ~15—30 GPa by repeated experiments [31,32,36,39-42,45]. There is no evidence for the well-known pentahedral intermediate state ($GeO_5$) found at 7—10 GPa [31,45,63].

Using the weak X-ray beam of 13-BM-D, the XAFS spectra at low pressures (≤6.4 GPa) are very similar, as are those at >20 GPa [35]. The major change in the Ge local environment occurs over the pressure range of 6–20 GPa [35]. The distinct change in Ge local environment at 5.4 GPa using the intense X-rays suggests a strong interaction of the X-ray photons with $GeO_2$ glassy network (lattice), which should be responsible the collapse of tetrahedral network in $GeO_2$ glass. This phenomenon has been never reported in network-forming glasses, to our best knowledge.

As shown in Fig. 2c, although the XAFS moduli $|\chi(R)|$ of X-ray induced octahedron (5.4 GPa) is close to that of pressure induced octahedron at 20.4 GPa in the range of 1.0–2.5 Å (without phase correction), X-ray induced octahedron shows significant modification at IRO distances in the range of 2.8–3.5 Å (without phase correction). In the light of the mechanism of permanent densified $GeO_2$ glasses, the densification largely comes from IRO modifications while small or no changes in the short-range order of the tetrahedra [64-67]. The arrow in Fig. 2c highlights remarkable changes at IRO distances, providing evidence for a different mechanism between X-ray and pressure induced densifications of $GeO_2$ glass.

Structural determination, including coordination distances and coordination numbers, is indispensable for a deeper comprehension of this X-ray induced phenomena. Fig.3 shows the evolution of the nearest Ge–O distance obtained from the first shell analysis at different X-ray irradiation experiments: run-1 (●), run-2 (▲) and quenched sample (○). The spot of run-2 was chosen at a fresh area (no X-ray illumination) 40 μm away from the centre of sample chamber (run-1) to crosscheck the observed X-ray induced structural change. All the XAFS parameters, amplitude, $E_0$, and the mean-square displacement (the Debye Waller factor) $\sigma^2$, were set free in the first shell fit. The obtained values of Ge–O distances by previous high-pressure XAFS experiments [35,39,40] are shown for comparison. The obtained Ge–O distances agree well with the values of tetrahedral $GeO_2$ glass from 0–2.7 GPa, but there is an abrupt elongation of Ge–O distances at the same pressure of 5.4 GPa upon X-ray irradiation. The average Ge–O distance at 5.4 GPa is 1.85±0.01 Å, which is longer than previous values for the octahedral glass at 20.4 GPa (Fig.3).

Determination of the Ge–O coordination number is necessary for characterizing the nature of X-ray induced phase transition in $GeO_2$ glass. As shown in Fig. 4, the obtained mean coordination number $N_{Ge}^{O}$ for $GeO_2$ glass exhibits a good agreement with the results of recent X-ray diffraction experiments [45] at pressures below 2.7 GPa. It can be noted that the Ge–O coordination number increases continuously close to a value of six for two independent runs, i.e., a lot of $GeO_6$ octahedral units has been formed upon X-ray irradiation at 5.4 GPa, confirming the spectroscopic evidences (Fig. 2) for the evolution from a continuous random network of corner shared tetrahedra to a dense octahedrally coordinated glass.

For the pressure-induced polyamorphs at low pressures (<6 GPa), $GeO_2$ glass first displays a decrease of intertetrahedral Ge-O-Ge angles and an increase of distortion of $GeO_4$ tetrahedra [35,45,68]. Above 6 GPa, compression takes places mostly through coordination changes with the formation of 5- or 6-fold Ge [31,35,36,45]. The coordination change is completed at 20 GPa, above which $GeO_2$ glass behaves as a fully octahedral glass with 6-fold Ge coordination [35,45]. The increasing density at 5.4 GPa, as represented by enhancement of absorption step (inset, Fig. 2a), can give important insights into the X-ray irradiation behavior of $GeO_2$ glass. It reflects the competing processes of increasing in coordination number (Fig. 4) and Ge–O bond lengthening in SRO distances (Fig. 3) upon X-ray irradiation. For pressure induced phase transition in $GeO_2$ glass, a first-order like phase transition in $GeO_2$ glass was proposed based on the rapid change of the Ge–O distances observed over a narrow pressure range between 7–9 GPa [36]. Here, we have observed a first-order sharp tetrahedral-octahedral phase transition at a fixed pressure (Fig.3 and Fig. 4). However, this X-ray induced changes of the Ge–O bond and coordination number keep heading on the octahedral state, but show no trapped intermediate state at pentahedral polyamorphs [31,45,63]. This example illustrates that the structural uncertainty due to X-ray irradiation should be considered for an accurate electronic and structural measurement using intense synchrotron X-rays.

It has been reported that octahedral form of $GeO_2$ glass can be induced thermally at low pressure of 5.3 GPa [43]. However, this thermal induced phase transition is very fast without any noticeable thermal delay[43]. The driving force behind this X-ray induced sluggish behavior of $GeO_2$ glass at high pressure is different from thermal effect (Fig. 1).

Fig.5 shows a comparison between the experimental spectrum of $GeO_2$ glass at 5.4 GPa after 45 min irradiation by X-ray (full circles) and the best-fit calculation (black line) based on the Ge–O and Ge–Ge cluster of rutile $GeO_2$ at 20.3 GPa [69] in $r$-space $GeO_2$ (Fig. 5a) and back-transformed $|\chi(R)|$ (1–3.1Å) in $q$-space (Fig. 5b). Within a 5-shell single-scattering model, we fixed the path degeneracies of rutile cluster, kept same variables for all paths, such as amplitude factors ($S_0^2$) and energy shift ($E_0$), while allow the path length $r_{Ge-O}$ and $r_{Ge-Ge}$, and mean square relative displacements (MSRD) $\sigma_{Ge-O}$ and $\sigma_{Ge-Ge}$ to vary, for a total of six structural variable parameters. The rutile-based structural modelling (black line) to the X-ray irradiated $GeO_2$ glass at 5.4 GPa gives an acceptable goodness-of-fit parameter, $R_w$, of 0.026. The quality of the agreement is a decent fit over the entire $r$ and $q$ range even for this simple six variable parameters for 5-shell single-scattering modelling (Fig. 5). The fit yields little modifications in all the Ge–O paths by shrinking -0.02±0.01Å but relatively large expansion in Ge–Ge paths by +0.07±0.02Å, i.e., smaller difference in SRO distances in contrast to that of IRO distances. This 5-shell single-scattering modelling results in two Ge–O paths at 1.79 Å (dashed line), four Ge–O paths at 1.84 Å (red line), two Ge–Ge paths at 2.89 Å (green line), four Ge–O paths at 3.25 Å (blue line), and eight Ge–Ge paths at 3.41 Å (cyan line). Same rutile-based modelling was carried out for the pressure induced octahedral $GeO_2$ glass at 20.4 GPa with a $R_w$ of 0.019, showing a same little shrinking in the Ge–O paths by -0.02±0.02Å but large shortening in Ge–Ge paths by -0.42±0.04Å. This results illustrate that the X-ray induced octahedral $GeO_2$ glass has some distinct difference from the pressure induced octahedral $GeO_2$ glass, although their SROs are nearly the same. This is common because some phase transitions triggered by light or X-ray irradiation exhibit novel transient phases not observed in thermal equilibrium [24,70,71].

In order to understand the observed X-ray induced phase transition in $GeO_2$ glass, we refer to the photoelectronic mechanism found in a variety of materials [4,9,10,15,19,20,72]. X-ray photon initially generates a core–hole excitation of Ge atom in the tetrahedral $GeO_4$ units of $GeO_2$ glass. Relaxation of core–hole excitation yields low-energy secondary electrons, holes, and fluorescence photons. When a photoelectron is released, there is a large relaxation of the glassy lattice around central Ge atom due to electron transfer between the Ge centers and the associated neighboring atoms. It is reported that such a strong electron-phonon coupling can drive the whole electron-lattice system towards a new quasi-equilibrium state [9,13,15,26,73-77]. Migration of these photoelectrons to the oxygen atom of neighboring $GeO_4$ unit, will form a coulombic interaction, pulling the $O^-$ atom toward the positive charge of $[GeO_4]^+$ unit. In general, the photoinduced phases are rather unstable and their lifetimes are very short, typically being between picoseconds and microseconds. Absorption of X-ray photons generates photoexcited $GeO_4$ units whose population grows with time until the percolation threshold is reached and the structural collapse of tetrahedral networking is triggered, leading to the formation of octahedral $GeO_6$ units. At low pressures, the coulombic interaction or structural relaxation is insufficient to overcome the energy barrier between $GeO_4$ and $GeO_6$ units. Certain pressure threshold is a prerequisite for the occurrence of X-ray induced tetrahedral-octahedral transition, as the case of thermal induced phase transition.

Because the energy of X-ray generated photoelectrons depends on the energy difference above the absorption edge. Recapture of the high-energy ('hot') X-ray photoelectron involves a substantial momentum transfer, and is thus not allowed [10]. The slow electrons generated by relatively low energy X-rays near the white line (Fig. 1) is readily caught by the oxygen atoms of neighboring tetrahedral $GeO_4$ units, leading to a transition from $GeO_4$ to $GeO_6$ units. For better understanding of the mechanism of X-ray induced structural phase transition in $GeO_2$, a fast XAFS measurement, e.g., quick-XAFS, is desired in future measurements.

## IV. Summary

We have observed an unusual X-ray induced first-order tetrahedral-octahedral phase transition in tetrahedral $GeO_2$ glass at a pressure of 5.4 GPa, based on *in situ* XAFS measurements using X-ray absorption spectroscopy in a nano-polycrystalline diamond anvil cell (NPD). The results show that structural phase transition in strong network forming $GeO_2$ glass under pressure can be triggered by X-ray irradiation in addition to the well-known external stimuli of pressure and temperature. The formation of dense materials induced by external stimuli such as X-ray, pressure, and temperature is crucial for understanding the planetary formation and differentiation. This result also shows that it is important to be aware of the nature of X-rays as an effective excitation source for a matter under study, especially for the emerging unprecedented highly brilliant synchrotron radiation, because some of the targeted metastable states under extreme conditions might become undetectable.

**Acknowledgment**

We would like to thank T. Lanzirotti, N. Lazarz, M. Rivers, S. Tkachev and P. DiDonna for the assistance with the experiments. The GSECARS sector of APS is supported by NSF (EAR-1128799) and DOE (DE-FG02-94ER14466). HKM was supported by NSF Grants EAR-1345112 and EAR-1447438. L. Huang acknowledges the support from NSF under grants DMR-1105238 and DMR-1255378.

## Figure Legends

FIGURE 1. X-ray irradiation effects on the raw unnormalized Ge $K$-edge XAFS of $GeO_2$ glass collected 2.7 GPa and 5.4 GPa, respectively. Left panel (a): overall spectra. Right panel (b): Expanded XANES region. Inset of Fig. 1a shows the evolution of XAFS absorption edge-step with X-ray irradiation as time increases.

FIGURE 2. The XAFS data of $GeO_2$ glass at 5.4 GPa in comparison with those measured for tetrahedral glass at 4.6 GPa and octahedral glass at 20.4 GPa, respectively, using a weak beam from a bending magnet source at 13-BM-D. A ratio of roughly $6 \times 10^5$ folds in X-ray brightness is achieved at 13-ID-E than that of 13-BM-D. (a) normalized Ge $K$-edge; (b) $k^2$-weighted XAFS spectra, $k^2\chi(k)$, in k-space; (c) XAFS Fourier transform, $|\chi(R)|$, in r-space; (d) Comparison of back-transformed $|\chi(R)|$ at 1–3 Å in $q$-space.

FIGURE 3. Evolution of the nearest Ge–O distance as X-ray irradiation time increases: run-1 (●), run-2 (▲) and quenched sample (○) coordinated models. Ge–O distances from previous high-pressure XAFS experiments [35,39,40] are shown for comparison.

FIGURE 4. X-ray induced variation of the mean coordination number $N_{Ge}^{O}$ for $GeO_2$ glass. The results are compared to the data of recent X-ray diffraction experiments [45].

FIGURE 5. (a) Fourier transformation modulus, $|\chi(R)|$, of $k^2\chi(k)$ for $GeO_2$ glass at 5.4 GPa after 45 min irradiation by X-ray (full circles). The rutile-based structural modelling (black line) yields two Ge–O paths at 1.79 Å (dashed line), four Ge–O paths at 1.84 Å (red line), two Ge–Ge paths at 2.89 Å (green line), four Ge–O paths at 3.25 Å (blue line), and eight Ge–Ge paths at 3.41 Å (cyan line). (b) Comparison of back-transformed $|\chi(R)|$ at 1–3.1 Å between experimental data (full circles) and best-fit calculation (black line) in $q$-space.

Fig. 1

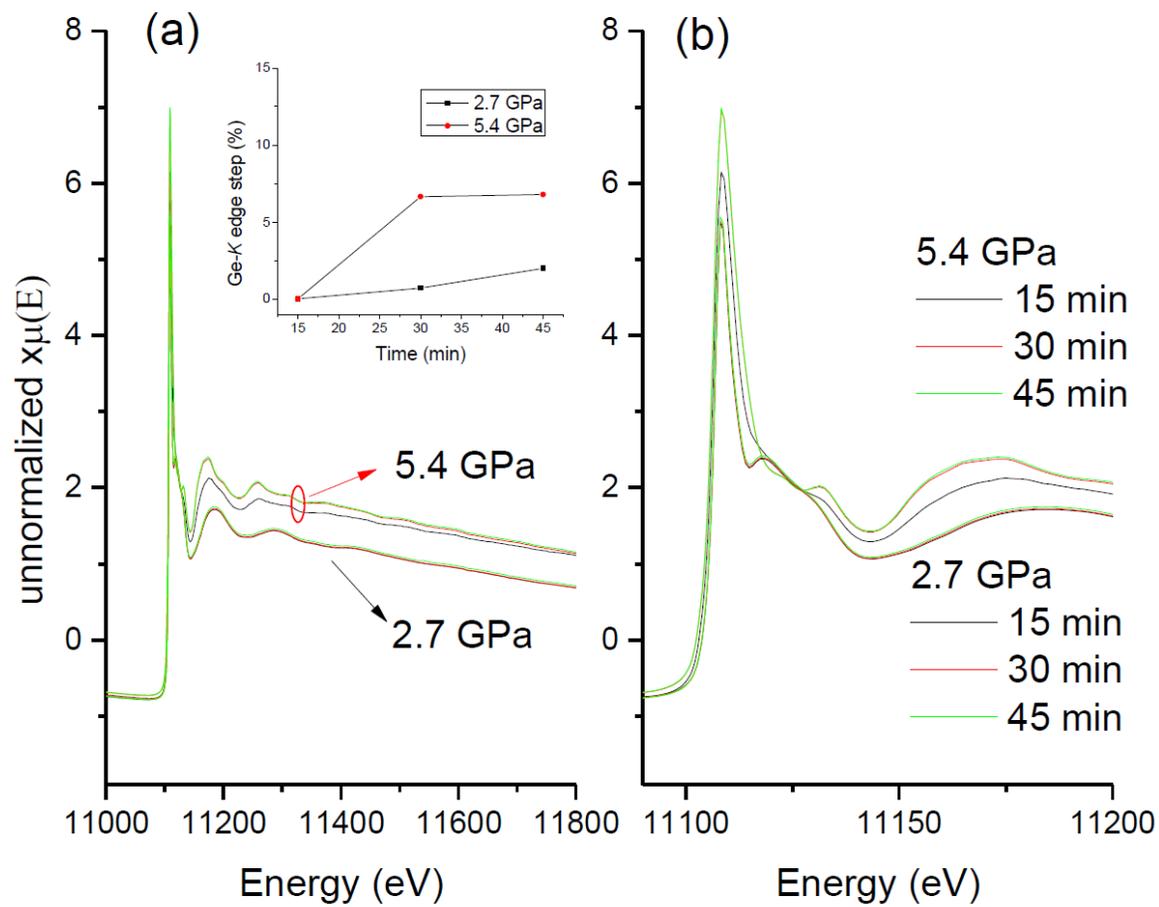

Fig. 2

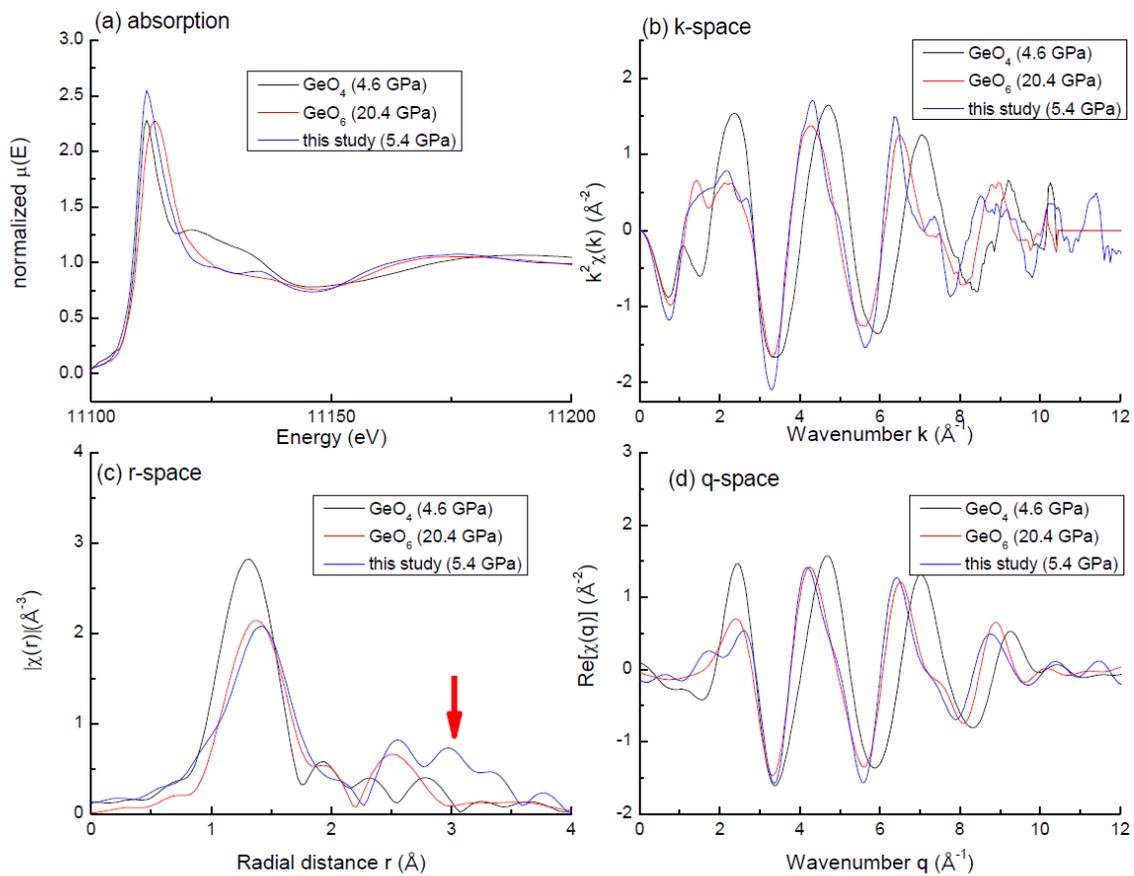

Figure 3

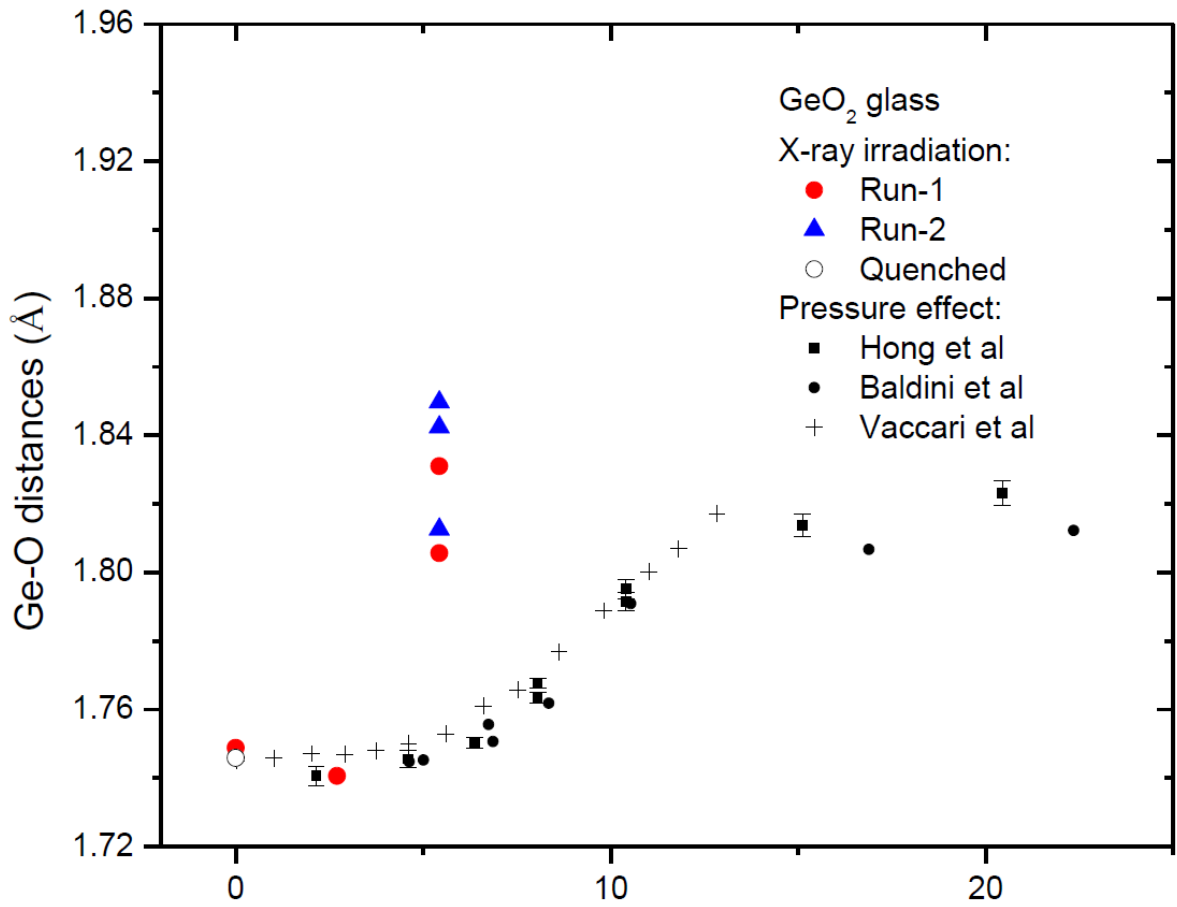

Figure 4

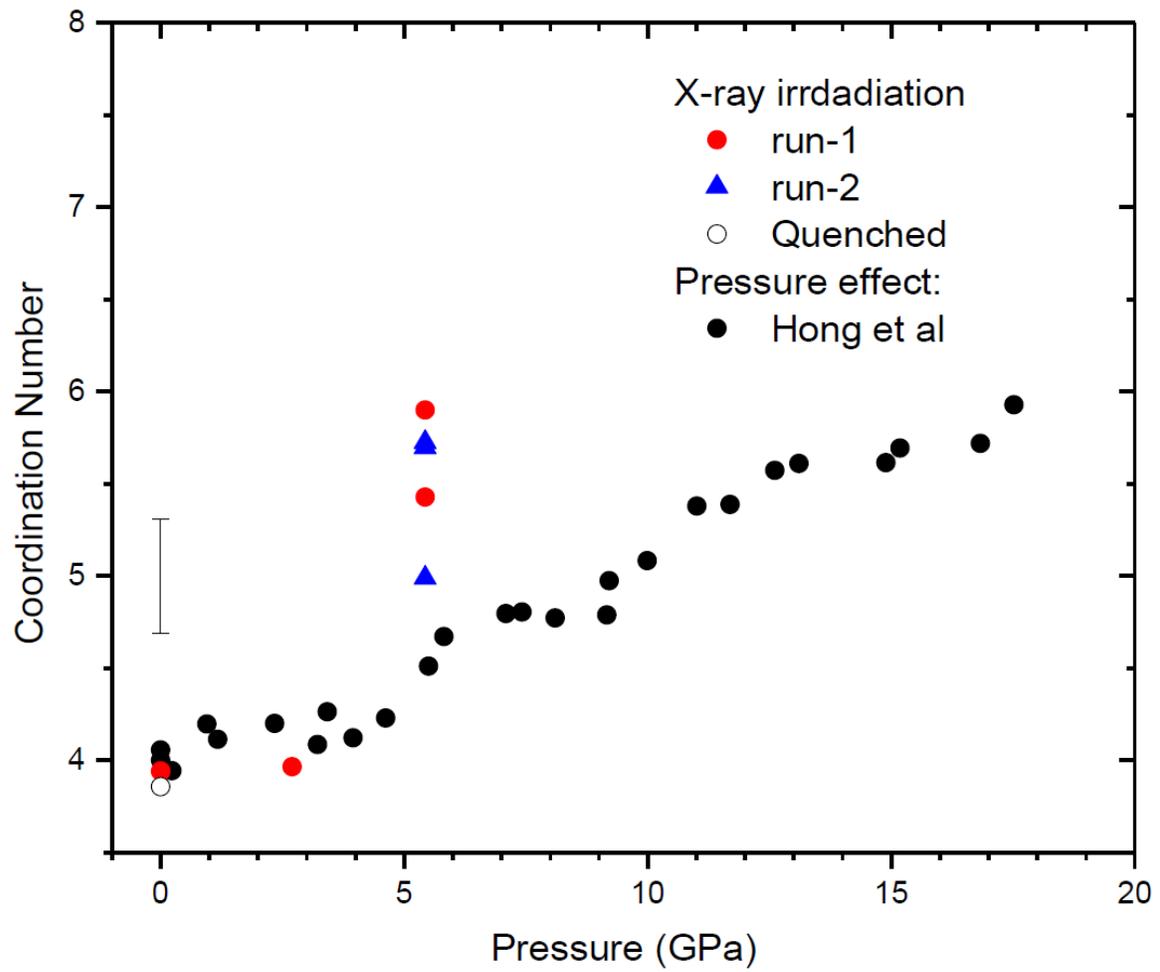

Figure 5

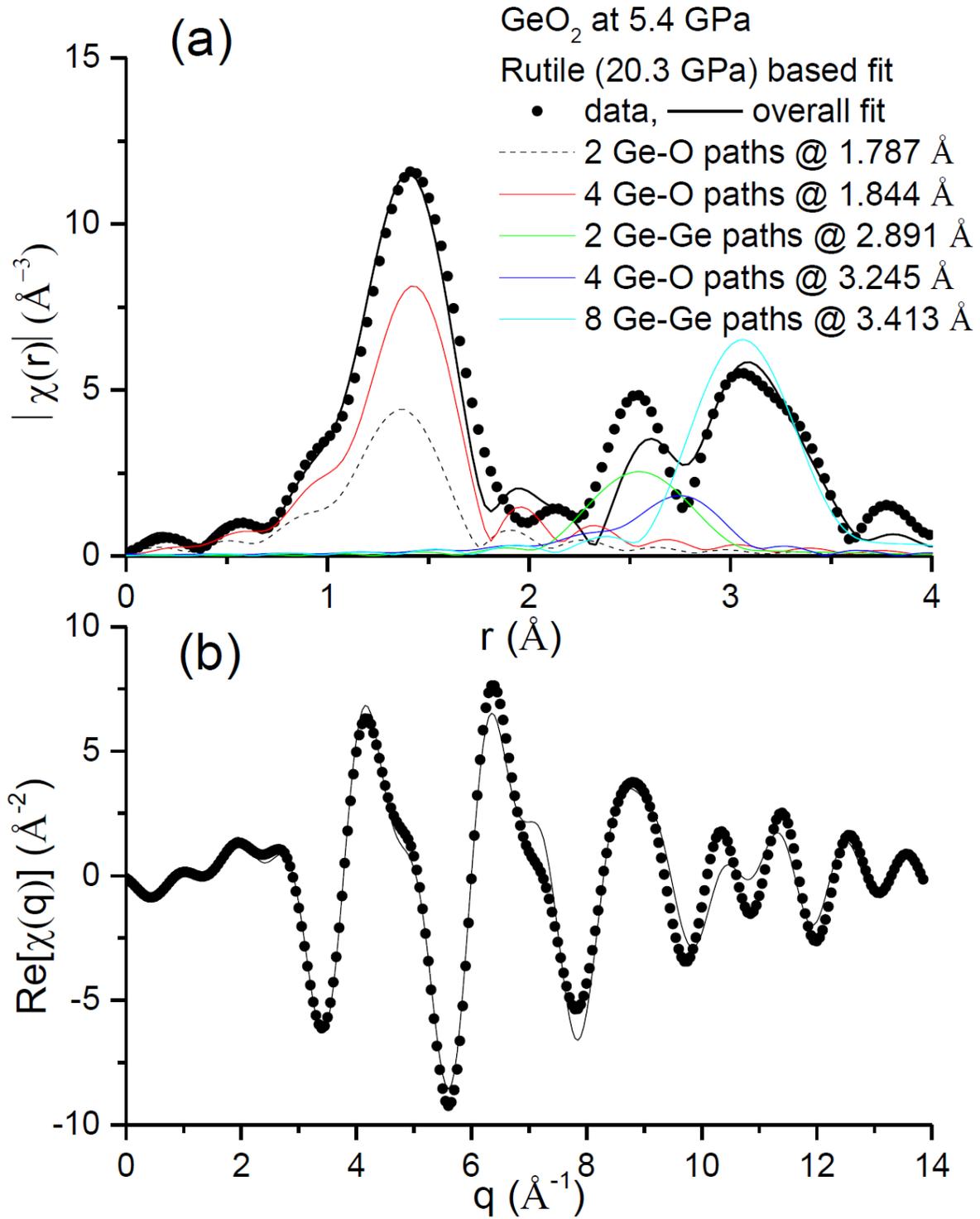

ss-Figures:

ss-FIGURE 1. XAFS Fourier transform, $|\chi(R)|$, for $GeO_2$ glass at 5.4 GPa as a function of X-ray irradiation time. Note the X-ray induced significant enhancement in the intermediate range order of three peaks (2.2–3Å).

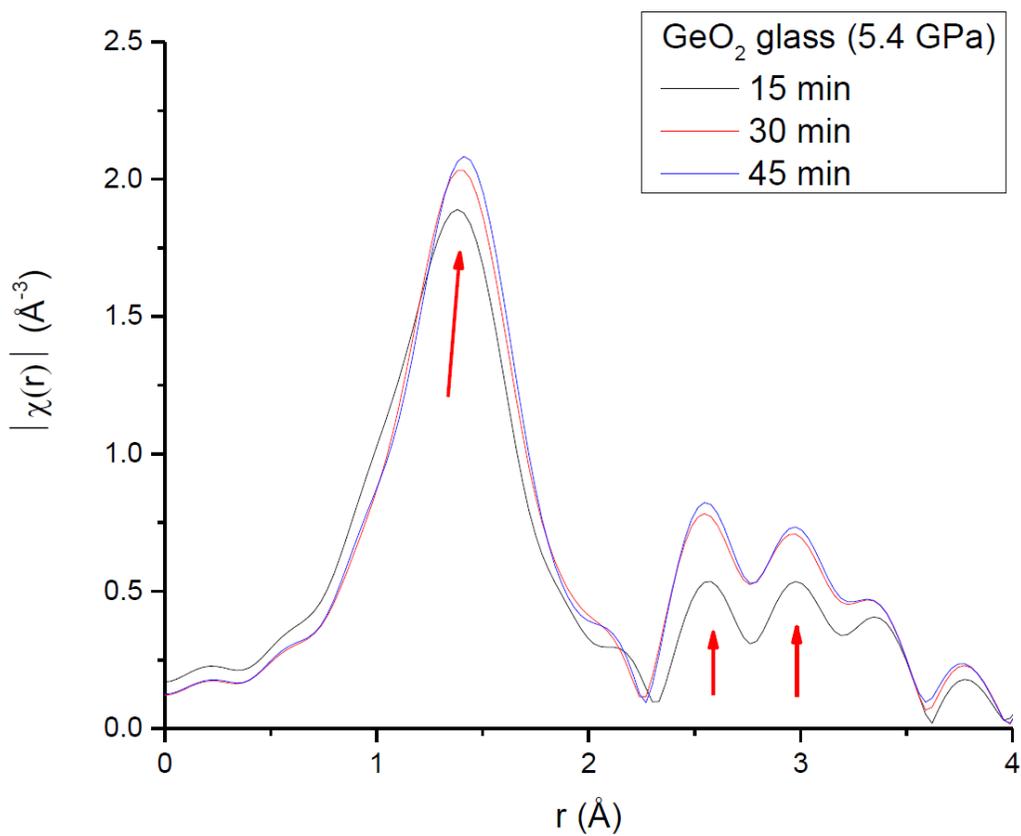

ss-FIGURE 2. (a) Normalized Ge $K$-edge XAFS of $GeO_2$ glass collected ambient pressure and quenched sample at 5.4 GPa, respectively. (b) $k^2$-weighted XAFS spectra, $k^2\chi(k)$, in k-space;(c) XAFS Fourier transform, $|\chi(R)|$, in r-space. The quenched sample doesn't display any detectable X-ray induced densification.

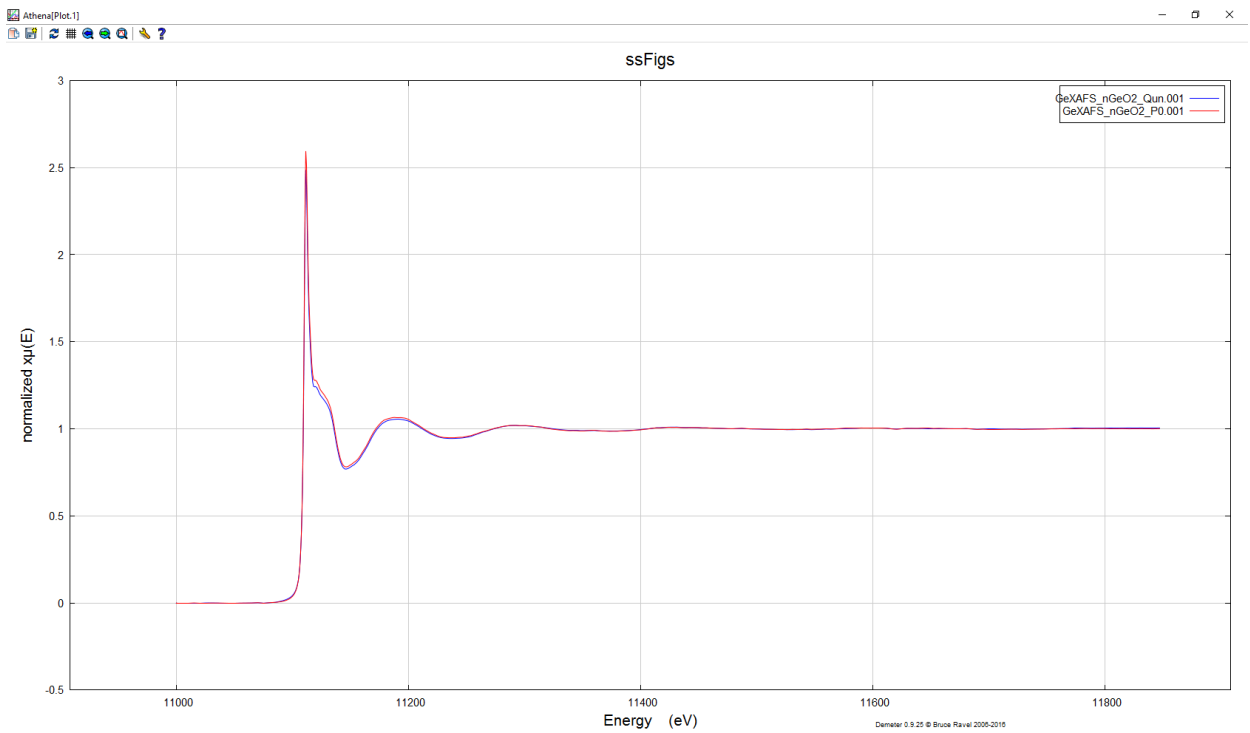

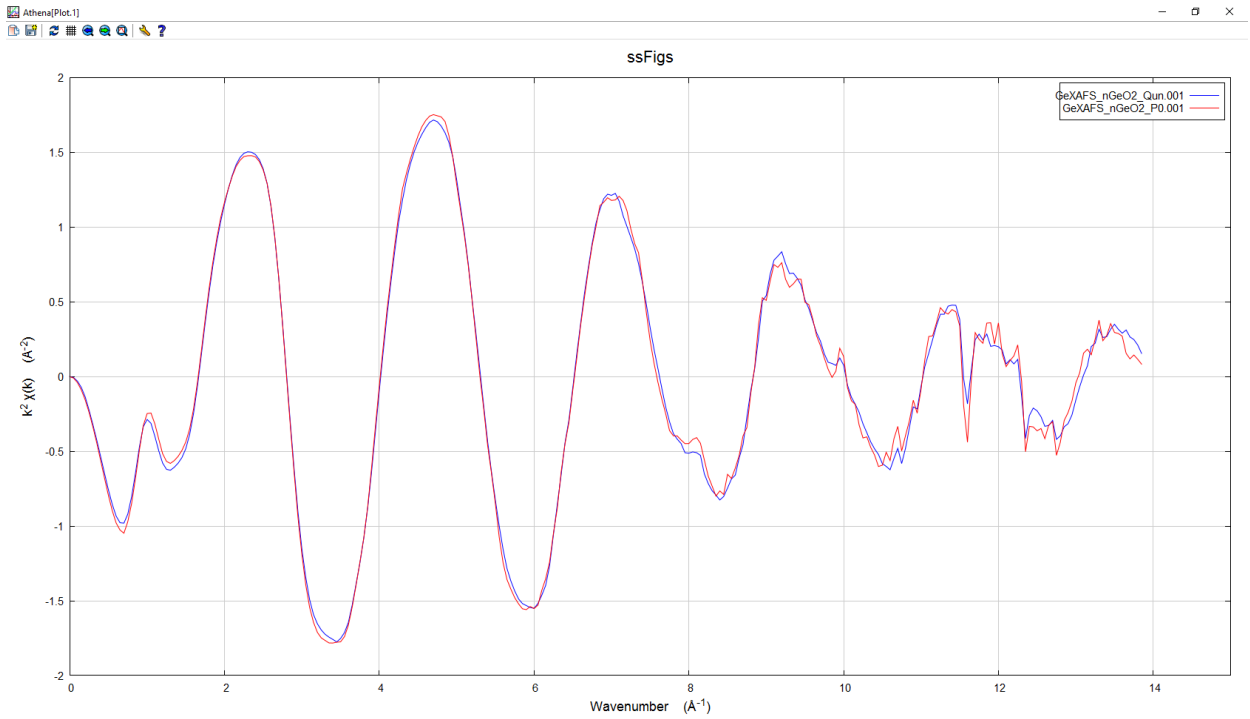

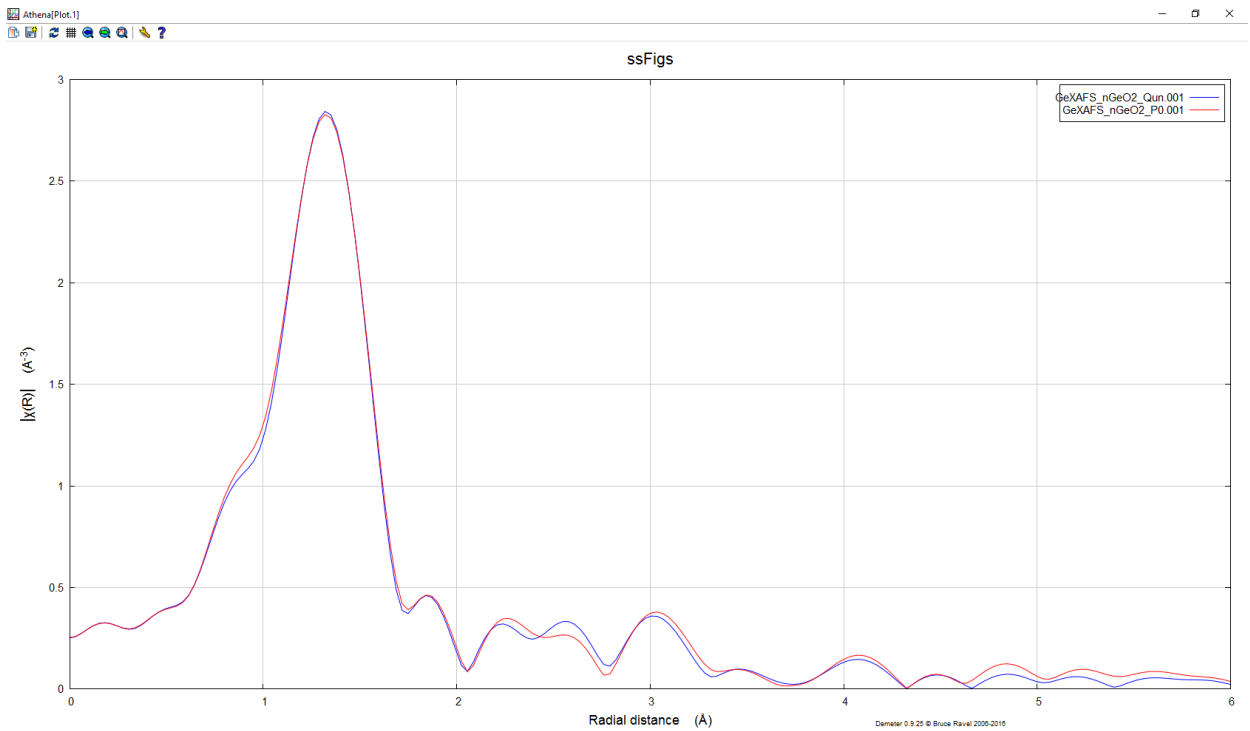

ss-FIGURE 3. X-ray induced damages on the surface of diamond anvils. (a) the nano-crystalline diamond anvil; (b) single crystalline diamond anvil.

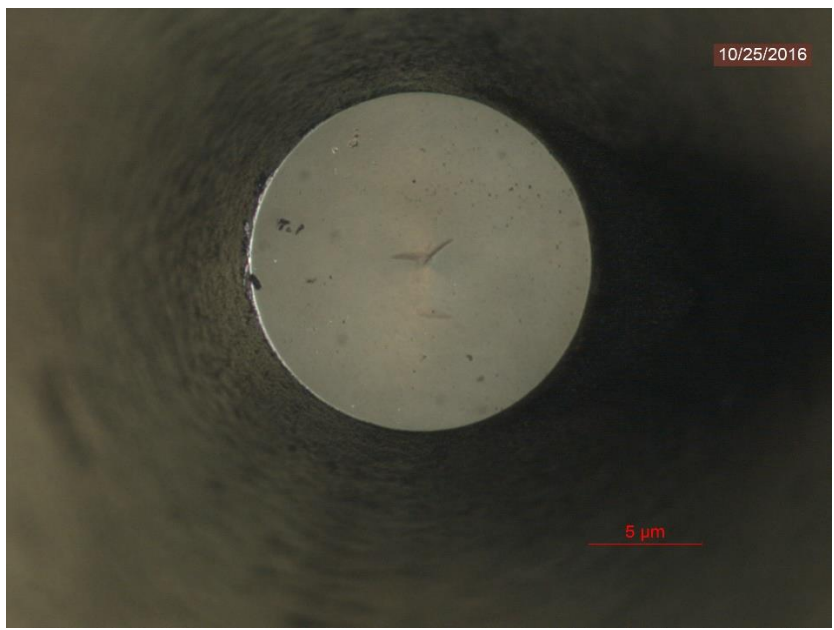

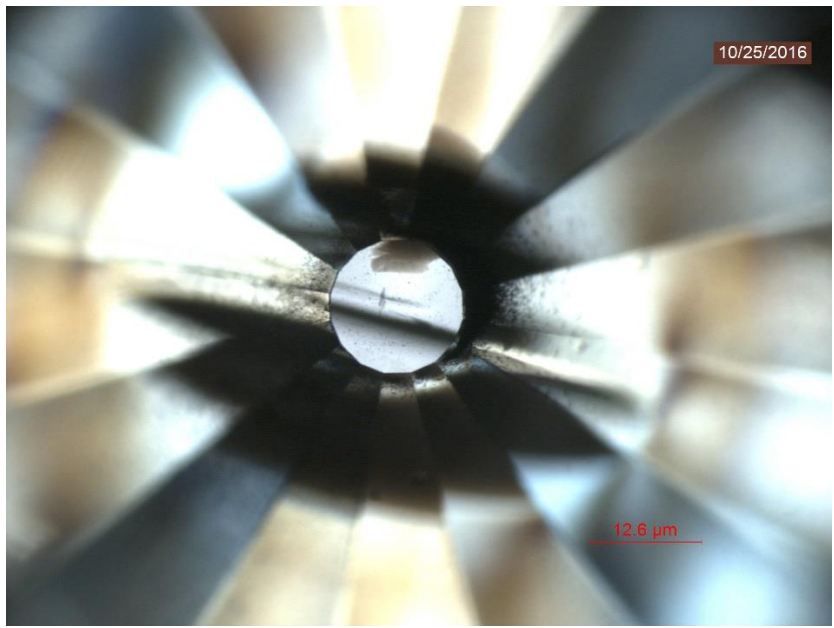